\documentclass[12pt]{iopart}

\usepackage{graphicx}

\begin{document}

\title[Is a soft nuclear equation of state incompatible with pulsar
data?]{Is a soft nuclear equation of state extracted from heavy-ion
data incompatible with pulsar data?}

\author{Irina Sagert$^1$, Mirjam Wietoska$^1$, J\"urgen
  Schaffner--Bielich$^1$\footnote{talk given at the international
    conference on 'Nuclear Physics in Astrophysics III', Dresden, March
    26--31, 2007}, Christian Sturm$^2$}

\address{$^1$Institut f\"ur Theoretische Physik/Astrophysik,
J. W. Goethe Universit\"at,
D-60438~Frankfurt am Main, Germany}

\address{$^2$Institut f\"ur Kernphysik,
J. W. Goethe Universit\"at,
D-60438~Frankfurt am Main, Germany}

\ead{schaffner@astro.uni-frankfurt.de}

\begin{abstract} 
  We discuss the recent constraints on the nuclear equation of state
  from pulsar mass measurements and from subthreshold production of
  kaons in heavy-ion collisions. While recent pulsar data points towards
  a hard equation of state, the analysis of the heavy-ion data allows
  only for soft equations of state. We resolve the apparent
  contradiction by considering the different density regimes probed. We
  argue that future measurements of global properties of low-mass pulsars
  can serve as an excellent cross-check to heavy-ion data.
\end{abstract}


\section{Introduction}


There has been a tremendous activity recently with regard to the
properties of dense nuclear matter in high-density astrophysics
of compact stars and in heavy-ion physics. The maximum mass of neutron
stars for example is intimately related to the underlying stiffness of
the nuclear equation of state (EoS), see e.g.~\cite{Lattimer:2006xb} and
references therein for a recent review.

More than 1600 pulsars, rotation-powered neutron stars, are presently known. New
timing measurements of binary pulsars point towards large masses and
correspondingly to a nuclear EoS usually considered to be
rather stiff. For the white dwarf--pulsar system J0751+1807 a mass of
$M=(2.1\pm 0.2) M_\odot$ ($1\sigma$) and $M=(1.6-2.5)M_\odot$ ($2\sigma$)
was found by Nice et al.~\cite{Nice:2005fi}. \"Ozel derives a mass limit
$M\geq (2.10 \pm 0.28) M_\odot$ from x-ray bursts of the neutron star EXO
0748--676 \cite{Ozel:2006km} stating that soft equations of state are
ruled out.

On the other hand, particle production and collective effects in
heavy-ion collisions are also considered to be a probe of the underlying
nuclear EoS. The subthreshold production of kaons as
measured by the KaoS collaboration (\cite{Sturm:2000dm} and
\cite{SengerNPA3}) were investigated in transport models (see
\cite{Fuchs:2000kp} and for a most recent reinvestigation see
\cite{Hartnack:2005tr} as well as \cite{Fuchs:2007vt}) coming to the
conclusions that the nuclear EoS should be considerably soft
above normal nuclear matter density.

We investigate in detail this apparent contradiction by considering
several families of nuclear equations of state, with regard to
compression modulus and symmetry energy, and their impact on the
properties of compact stars taking into account the recent data on kaon
production in heavy ion collisions. We note that a combined analysis of
other heavy-ion data (not on kaon production data) and compact stars has
been performed in \cite{Klahn:2006ir}. In particular, the role of the
density region relevant for subthreshold production and the related
Schr\"odinger equivalent potentials are delineated. In addition, the
importance of hyperons in dense neutron star matter with regard to known
hypernuclear properties are contrasted with pulsar mass measurements.
Also, the influence of a possible quark phase on the mass-radius
relation of compact stars is discussed, in particular in response to the
recent analysis by \"Ozel (see \cite{Alford:2006vz}).


\section{The nuclear equation of state and high-density 
astrophysics}


The nuclear EoS serves as a crucial input for modelling astrophysical
systems with extreme matter densities as core-collapse supernovae (for a
review see \cite{Janka:2006fh}), neutron star mergers (see
e.g.~\cite{Rosswog:2001fh,Oechslin:2006uk}), proto-neutron star
evolution \cite{Pons:1998mm} and cold neutron stars \cite{Weber:2004kj}.
The required density range is huge, spanning densities from about
$10^{-10} n_0$ to up to $10 n_0$, where $n_0=0.15$ fm$^{-3}$ stands for
the normal nuclear matter density (the low-density region is of
particular importance for dynamical systems, core-collapse supernovae
and neutron star mergers).  The temperature scale of interest spans from
the keV range up to 50 MeV, for neutron star mergers even larger
temperatures might be possible which depends sensitively on the
stiffness of the nuclear EoS. The nuclear equations of state presently
used in most dynamical simulations are the ones of Lattimer and Swesty
\cite{Lattimer:1991nc} and Shen et al.~\cite{Shen:1998gq}. On the other
hand, there are dozens of nuclear equations of state used for describing
proto-neutron stars and cold neutron stars, see e.g.~\cite{Weber:2007ch}
and references therein.

Stars with a zero-age main sequence mass of more than 8 solar masses end
in a core-collapse supernova (type II).  A new generation of simulation
codes has been developed during the last few years including first steps
towards a full 3D-treatment and Boltzmann neutrino transport.  These
improved models of stellar core collapse have found no explosions
initially and the question on the missing physic ingredient was raised
(see \cite{Buras:2003sn}). We quote from this reference: ``\dots the
models do not explode. This suggests missing physics, possibly with
respect to the nuclear equation of state \dots''. (After the conference
the situations has changed substantially. In particular, the newest
simulation found indeed that the models do explode. A 15 solar mass
progenitor star finally exploded after about 600ms due to the standing
accretion shock instability, dubbed SASI, see \cite{Janka:2007yu}.)

The remnants of core-collapse supernovae (type II and type Ib/Ic) are
neutron stars, compact massive objects with typical masses of
$(1-2) M_\odot$ and radii of about 10~km. The link between pulsars,
rotation-powered neutron stars, and supernova remnants is most evident for the
crab pulsar and the crab nebula, the supernova remnant of the historical
supernova of 1054 A.D\@. Nowadays, more than 1600 pulsars are known, the
number is increasingly steadily due to recent detailed radio pulsar
scans. However, the list of well determined pulsar masses is restricted
to a few dozens, to binary systems (for a most recent compilation from
radio observations see \cite{Stairs:2006yr}). The best determined
masses from relativistic double neutron star systems are between $1.44
M_\odot$ for B1913+16 (the Hulse-Taylor pulsar) and $(1.18\pm
0.02)M_\odot$ for the pulsar J1756--2251 \cite{Faulkner:2005}.

Of enormous interest for the nuclear EoS is the report on
the measurement of massive neutron stars in pulsar--white dwarf systems,
in particular for the pulsar J0751+1807 the mass constraints $M=(2.1\pm
0.2) M_\odot$ ($1\sigma$) and $M\geq 1.6 M_\odot$ ($2\sigma$) have been
reported \cite{Nice:2005fi}. The larger the mass pulsar limit the more can
the nuclear EoS be constrained at high densities. Also measurements
of neutron star radii will help to learn more about the physics of the
interior of neutron stars.  The isolated neutron star RXJ 1856, the
closest known neutron star, emits a rather perfect x-ray black-body
spectrum. However, combined with the optical flux measurements by the
Hubble Space Telescope, a two temperature fit is needed. A small soft
temperature implies a rather large radius so that the optical flux is
right giving a conservative lower limit for the radiation radius of
$R_\infty = 16.5$ km (d/117 pc) \cite{Trumper:2003we}. Note, that such a
large radiation radius rules out most of the nuclear equations of state
presently discussed in the literature, not only for quark matter but
also for ordinary nuclear matter composed of neutrons, protons and
leptons only.

The X-Ray burster EXO 0748--676 has been recently reanalysed by \"Ozel
\cite{Ozel:2006km} who concludes that the mass and radius are constrained
by $M\geq (2.10\pm 0.28) M_\odot$ and $R\geq (13.8 \pm 1.8)$~km. One important
ingredient is the redshift extracted from spectral lines
\cite{Cottam:2002cu}. Both, mass and radius, are quite large so that
very soft nuclear equations of state, could be ruled out. The appearance
of new degrees of freedom in nuclear matter, as hyperons or kaon
condensates, can considerably soften the nuclear EoS as
the new degrees of freedom fill new low-lying Fermi levels. However,
quark matter is a completely new phase with entirely new properties, so
that the quark matter EoS can be rather stiff. Hence, pure
quark stars as well as hybrid stars, neutron stars with nuclear and
quark matter components, can be well within the above mass and radius
limits \cite{Alford:2006vz}. 

In any case, there seems to be growing evidence from pulsar data that
the nuclear EoS should be hard at high densities.


\section{The nuclear equation of state and 
heavy-ion physics}


On the other hand, kaon production at subthreshold energies in heavy-ion
collisions arrives at the conclusion that the nuclear EoS
is rather soft (see also the talks by Peter Senger and Christian Fuchs
\cite{SengerNPA3,Fuchs:2000kp})!

Nucleus-nucleus collisions at relativistic energies offer the unique
possibility to study experimentally the properties of dense nuclear
matter in the laboratory. In a reaction between two heavy nuclei at beam
energies around 1 AGeV nuclear matter will be compressed up to three times
saturation density. Therefore, heavy-ion experiments provide the
possibility to obtain information on the nuclear EoS at
high baryon densities.

In order to obtain information on the high-density phase of a
nucleus-nucleus collision it is necessary to measure particles which are
created predominantly in this phase. At the energy domain of the SIS at
GSI Darmstadt $K^{+}$ mesons seem to be promising candidates due to
their production mechanism and their long mean free path in nuclear
matter. The propagation of $K^+$ mesons in nuclear matter is
characterised by the absence of absorption (as they contain an
antistrange quark) and hence they are almost undisturbed messengers.

The production of $K^+$ mesons requires multiple nucleon-nucleon
collisions or secondary collisions such as $\pi N \to K^{+} \Lambda$ at
subthreshold beam energies ($E_{\rm beam} = 1.58$~GeV for $NN \to K^{+}
\Lambda N$). Since these multi-step processes occur preferentially at
high baryon densities $K^{+}$ mesons are expected to be produced mostly
during the high density phase of the reaction. Hence, the $K^+$ yield
depends sensitively on the maximum baryon densities reached
within the reactions which is determined by the stiffness of the nuclear
EoS. Therefore, $K^+$ mesons are expected to be well
suited to obtain information on the nuclear EoS at high
baryon densities.

Early transport calculations predicted that the $K^+$ yield from Au+Au
collisions at subthreshold energies will be enhanced by a factor of
about two if a soft rather than a hard EoS is assumed
\cite{Aichelin:1986ss}. More recent calculations (IQMD
\cite{Hartnack:1993bq}, RQMD \cite{Fuchs:2000kp}) were performed with
two values for the compression modulus of 200~MeV and 380~MeV
which correspond to a ``soft'' or ``hard'' nuclear EoS,
respectively. The transport models take into account a repulsive
$K^{+}N$ potential and use momentum-dependent Skyrme forces to determine
the compressional energy per nucleon (i.e.\ the energy stored in
compression) as a function of the baryon density.

The assumed repulsive $K^{+}N$ potential depends nearly linear on the
baryonic density \cite{SMB97} and thus reduces the $K^+$ yield
accordingly. On the other hand, at subthreshold beam energies the
$K^{+}$ mesons are created in secondary collisions involving two or more
particles and hence the production of $K^+$ mesons depends at least
quadratically on the density. To disentangle these two competing effects
the KaoS Collaboration \cite{Senger:1992gz} has studied $K^+$ production
in a very light ($^{12}$C+$^{12}$C) and a heavy collision system
($^{197}$Au+$^{197}$Au) at different beam energies near threshold. Due
to pile-up of nucleons the maximum baryonic density in Au+Au reactions
is significantly higher than in the light collision system C+C.
Moreover, the maximum baryonic density reached in Au+Au reactions
depends on the compression modulus of nuclear matter whereas in C+C
collisions this dependence is very weak \cite{Fuchs:2000kp}.
Additionally, the use of the Au/C ratio provides the advantage that
uncertainties of the experiment (beam normalisation etc.) as well as of
the transport model calculations (elementary cross sections etc.) cancel
partly.

The ratio of the $K^{+}$ production excitation functions for Au+Au and
C+C reactions obtained by the KaoS Collaboration increases with
decreasing beam energy \cite{Sturm:2000dm}. The comparison with
transport model calculations demonstrates clearly that only the
calculations based on a soft nuclear EoS reproduce the
trend of the experimental data~\cite{Fuchs:2000kp}.


\section{Probing the nuclear equation of state}


In this section, we examine the constraints from astrophysical and
heavy-ion data using an empirical nucleon-nucleon interaction.  As
ansatz for the energy per particle we take the standard Skyrme-type
parameterisation, which is also used in the above mentioned analysis of
the heavy-ion data: 
\begin{equation}
\epsilon/n = m_N + E^{kin}_0  + \frac{A}{2} \cdot u 
+ \frac{B}{\sigma + 1} u^\sigma + S_0 \cdot u \cdot
\left(\frac{n_n-n_p}{n}\right)^2
\end{equation}
where $u=n/n_0$. The parameters $A$, $B$, $\sigma$ are fixed by the
nuclear matter properties, i.e.\ by normal nuclear matter density $n_0$,
the binding energy $E/A$, the compression modulus $K$, and the asymmetry
term by the asymmetry energy $S_0$ at $n_0$.  The pressure is determined
by the thermodynamic relation $P = n^2 d(\epsilon/n)/dn$, which fixes
the EoS underlying the transport model calculations of
\cite{Fuchs:2000kp,Hartnack:2005tr}. We note that the equation of state
can become acausal for $\sigma>2$. With this EoS at hand, we can check
whether or not low compressibilities are ruled out by neutron star mass
measurements or not.

\begin{figure}
\centerline{\includegraphics[angle=-90,width=1.0\textwidth]{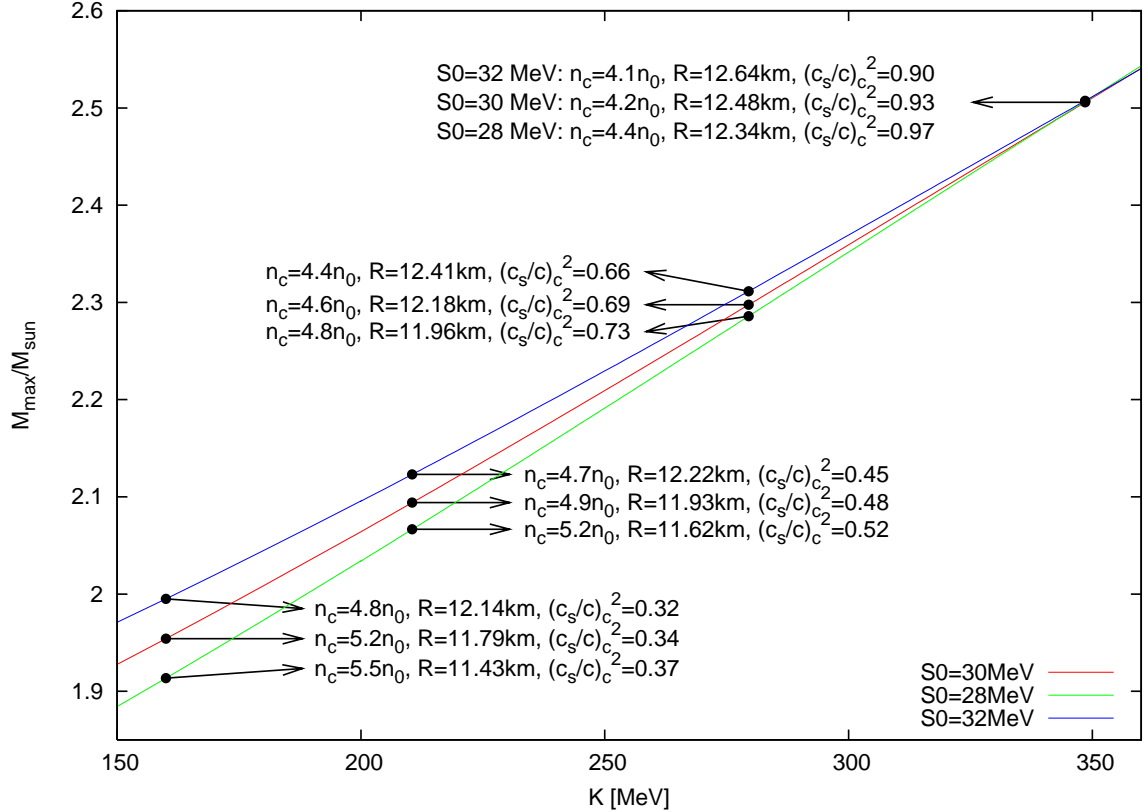}}
\caption{The maximum mass as a function of the compression modulus $K$
  for nucleons and electrons only in $\beta$-equilibrium, choosing the
  symmetry energy to be $S_0=28,30,$ and 32 MeV. For some points on the
  curve the central density, the radius, and the maximum speed of sound
  squared are denoted. A compression modulus of $K\geq 190$ MeV is
  sufficient to arrive at a maximum mass above 2$M_\odot$.}
\label{fig:maxmass}
\end{figure}

Figure~\ref{fig:maxmass} shows the results for the maximum mass
configuration of neutron stars in dependence of the compression
modulus $K$. Plotted is the maximum mass, at certain points the
corresponding radius, maximum speed of sound, and the maximum density
are denoted in addition. The maximum central density reached is $n_c
\approx 5n_0$ for $K=200$ MeV. There is a small variation of the radius
on the compression modulus $K$ and the EoS stays causal,
i.e.\ $c_s^2\leq 1$, up to $K=350$ MeV. We also find that
there is only a slight dependence on the symmetry energy $S_0$, up to
$\Delta M=\pm 0.05M_\odot$ for low values of the compression modulus for
$S_0=(28-32)$~MeV. Most importantly, however, the maximum mass can be
larger than two solar masses, $M\geq 2M_\odot$, for a a compression
modulus above $K\approx 190$~MeV. Hence, even 'soft' equations of state
in terms of a small compression modulus can give large neutron star
masses!

The above analysis has been repeated using relativistic mean-field (RMF)
models. It is important to realize that the essential input to transport
models is the non-relativistic Schr\"odinger equivalent potential, which
determines the propagation of the nucleons in the transport simulation,
not the EoS directly (this is not important for the non-relativistic
model used above). We derive a non-relativistic nucleon potential from
the Skyrme parameterisation corresponding to $K=200$~MeV.  We compare
this nucleon potential to that of RMF parameterisations including a
possible vector self-interaction term and allow for only those parameter
sets which have a Schr\"odinger equivalent potential compatible with the
one from the Skyrme parameterisation. We find indeed RMF parameter sets
fulfilling that constraint which have maximum neutron star masses around
two solar masses if only nucleons and leptons are taken into account.
Hence, soft (non-relativistic) potentials can be also compatible with
pulsar masses! A most recent analysis using relativistic
Brueckner-Hartree-Fock models arrives at a similar conclusion, see the
contribution by C.~Fuchs \cite{Fuchs:2007vt}.

However, the maximum density reached is about $5n_0$. At such large
densities, various models describing nuclei as well as hypernuclear data
find consistently that hyperons are present already at $n=2n_0$ in
neutron star matter, see e.g.\ the discussion in
\cite{SchaffnerBielich:2002ki}. Also, kaon condensates as well as quark
matter could be present in the core of neutron stars at such large
densities. Hyperons drastically reduces the maximum mass possible (we
refer to \cite{GM91} for a particular nice plot demonstrating this
effect). Hyperons are a new degree of freedom, therefore the overall
reduced pressure does not allow for large neutron star masses. Indeed,
we find that the inclusion of hyperons reduces the maximum masses to
about $1.75M_\odot$ and below in RMF models. For the RMF parameter sets
with sufficiently soft Schr\"odinger equivalent potentials compatible
with the KaoS data masses below $1.6 M_\odot$ (the lower limit from
J0751+1807) are found, although masses above $1.44 M_\odot$ are
possible. It is remarkable that the presence of hyperons have such an
effect on the maximum mass, although they are only present above $2n_0$.

Hence, we conclude that the relevant density region probed by the
maximum mass and mass limits from pulsar data is most likely above
$2n_0$. So the high density EoS is crucial for the maximum
mass which is controlled by unknown physics (hyperons, kaon
condensation, quark matter). Kaon production data from heavy-ion
experiments on the other hand are sensitive to moderate densities of
around $2n_0$ only. Different density regions are probed so that
present heavy-ion data and pulsar mass limits can not be compared
directly. However, this finding immediately points towards a real
comparison: to check for the global properties of low-mass neutron stars
with moderate maximum densities! Within the empirical nucleon-nucleon
interaction it turns out that a $1.2M_\odot$ neutron star probes at
maximum the EoS up to $n\approx 2n_0$, the right maximum density
so that exotic states are likely to be not present. The lowest measured
mass of a pulsar is the one of J1756--2251 with $(1.18\pm 0.02)M_\odot$
\cite{Faulkner:2005}. A measurement of the radius of such a low-mass
neutron star would allow for a more consistent cross-check between the
nuclear EoS derived from heavy-ion experiments and pulsar
observations.


\section{Summary}


We have demonstrated that a soft nuclear EoS as extracted from kaon
production data is not in contradiction with recent pulsar data. The EoS
above $n\approx 2n_0$ determines the maximum mass of neutron stars which
is controlled by unknown high-density physics (hyperons, kaon
condensation, quarks). Neutron stars with hyperons have rather low
maximum masses, so that he presence of hyperons makes it substantially
more difficult to comply with heavy-ion and pulsar data. Properties of
low-mass neutron stars ($M\approx 1.2M_\odot$), on the other hand, are
entirely controlled by the EoS up to only $n\approx 2n_0$.  Measurements
of the radius of low-mass pulsars open therefore the exciting
opportunity for a consistent cross-check between heavy-ion and pulsar
data on the properties of the nuclear equation of state at moderately
high densities.


\ack
This work is supported in part by the Gesellschaft f\"ur
Schwerionenforschung mbH, Darmstadt, Germany. Irina Sagert gratefully
acknowledges support from the Frankfurt Institute for Advanced Studies
and the Helmholtz Research School for Quark Matter Studies.

\section*{References}

\bibliographystyle{revtex}
\bibliography{all,literat,npa3_proc}

\end{document}